\documentstyle[12pt,fleqn]{article}
\def\dg{\dagger}

\def\uqn{su_q (n) }
\def\uqn'{u_q (n+1) }
\def\la{\langle}
\def\ra{\rangle}

\def\xb{{\bf x}}     \def\yb{{\bf y}}  

\def\T{\Theta }   \def\t{\theta }
\def\a{\alpha }   \def\b{\beta }  \def\g{\gamma }
\def\d{\delta }  \def\O{\Omega }  \def\D{\Delta }  \def\e{\epsilon }
\def\s{\sigma }  \def\S{\Sigma }

\def\beq{\begin{equation}}
\def\eeq{\end{equation}}

\def\vac{\vert 0\rangle }

\begin{document}

\begin{flushright}
 \mbox{hep-ph/9912222\hspace{6mm}}
\end{flushright}
\bigskip
\bigskip

{\Large
  \centerline{\bf Masses of decuplet baryons treated within}
    \vspace{0.1mm}
  \centerline {\bf anyonic realization of the $q$-algebras
$U_q({\rm su}_N)$ } }

\vspace{8mm}
\centerline{\sl A.M.~GAVRILIK,\ \ N.Z.~IORGOV}
\medskip
\centerline{\it Bogolyubov Institute for Theoretical Physics,}
    \vspace{0.1mm}
  \centerline {\it  03143 Kiev-143, Ukraine}

\vspace {9mm}
\begin{abstract}
\noindent
In the approach to hadronic flavour symmetries 
based on the
$q$-algebras $U_q(su_N)$ and proved to be realistic,
the known construction of $U_q(su_N)$ in terms of anyonic
oscillators residing on 2$d$ lattice is utilized.
Anyonic Fock-like realization of basis state vectors
is given for baryons $(3/2)^+$ from the ${\bf 10}$-plet of
$U_q(su_3)$ embedded, via
${\bf\sl 20}$-plet of $U_q(su_4)$, into the "dynamical"
representation $[4000]$ of $U_q(su_5)$.
  Within the anyonic picture, we reobtain the universal
$q$-deformed decuplet mass relation
$M_\O -M_{\Xi^*}+M_{\S^*}-M_{\D}=[2]_q(M_{\Xi^*}-M_{\S^*})$,
where $[2]_q=q+q^{-1}=2\cos{\t}$. Consistency with data for
baryon masses requires $\t\simeq\frac{\pi}{14}$.
As a result, anyons with anyonic statistics parameter
$\nu = \frac{1}{14}$
can be put into correspondence, at least formally, with
the constituent quarks of decuplet baryons.
\end{abstract}

\baselineskip 18pt

\vspace {6mm}
\noindent {\bf 1. Introduction}

\vspace {2mm}
Quantum groups and quantum (or $q$-deformed)
algebras $su_q(N)\equiv U_q(su_N)$
introduced more than a decade ago \cite{DFW,J}
remain to be a subject of intensive study.
Interesting and very important for further applications aspect of the
$q$-algebras consists in a variety of their possible
realizations. Among these, there exist realizations in
terms of $q$-deformed bosonic oscillators \cite{B-M,Ha}, $q$-fermionic
oscillators as well as mixed ($q$-bose {\it and} $q$-fermi)
ones \cite{Ha,FSS}.
Few years ago, it was shown by Lerda and Sciuto \cite{LS}
that the $q$-algebra $U_q(su_2)$ admits a realization in terms of
two modes of so-called anyonic oscillators - certain {\it non-local}
two-dimensional objects defined on a $2 d$ square lattice.
Shortly after that,
their results were extended \cite{CM}-\cite{FMS} to the
higher rank $U_q(sl_N)$ algebras and, moreover, to the $q$-analogs
of all semisimple Lie algebras from the classical series
$A_r $,  $B_r  $, $C_r$ and $D_r$.

On the other hand, $q$-algebras find their phenomenological
applications in (sub-)nuclear physics, e.g., in modelling
rotational spectra of deformed nuclei \cite{Iwa}, in describing
some properties of hadrons, see \cite{Kib,G97} for relevant
references. Concerning the use of higher rank $q$-algebras
for deriving generalized hadron mass relations, let us mention
the approach  based on replacing \cite{G93} usual unitary groups
of hadronic $(N\equiv N_f)$-flavor symmetries, or their Lie
algebras, by the corresponding quantum algebras $U_q(su_N)$.
Such replacement leads to a number of interesting implications
\cite{G97}-\cite{GKT}. In conjunction with this, the attempt
to utilize the possibility suggested by anyonic construction
of the $q$-algebras, within just mentioned approach to hadron
mass relations, seems quite natural.

Our goal in this note is to demonstrate that the $U_q(su_N)$ based
results concerning baryon masses and their mass sum rules,
at least for the decuplet case \cite{GKT}, can be rederived
in the framework of anyonic picture if the masses are defined
in such way that one gets rid of anyons' nonlocality in
final results. As a consequence, it appears that constituent
quarks within the present approach can be formally treated
as anyons of definite statistics $\nu $.
A comparison of the obtained $U_q(su_N)$ mass relation for
decuplet baryons with data for empirical masses shows:
one has to fix the anyonic statistics parameter
(directly linked to the deformation strength) as $\nu ={1\over 14}$.

\vspace {4mm}
\noindent {\bf 2. Anyonic oscillators on $2 d$ square lattice}

\vspace {3mm}
In this section we recapitulate necessary minimum of details
concerning $d=2$ lattice angle function, disorder operators,
and anyonic oscillators (see \cite{LS}-\cite{FMS}).
Let $\O$ be a $2 d$ square lattice with the spacing
$a\! =\! 1$.
On this lattice we consider a set of $N$ species (sorts) of fermions
$c_i ({\bf x})$,\ \ $i=1,...,N$,\ \ ${\bf x}\in\Omega$,\ which
satisfy the following standard anti\-commutation relations:
\beq   \label {1}
\{ c_i ({\bf x}), c_j ({\bf y})\} =
\{ c_i^\dg ({\bf x}), c_j^\dg ({\bf y})\} = 0,
\eeq
\beq         \label {2}
\{ c_i ({\bf x}), c_j^\dg
({\bf y})\} =\d_{ij}\ \d ({\bf x}, {\bf y}) .
\eeq
Here $\d({\bf x}, {\bf y})$\ is
the conventional
lattice $\d$-function: $\delta ({\bf x}, {\bf y}) = 1$\ if
${\bf x}={\bf y}$\ and vanishes if ${\bf x}\neq {\bf y}$.

We use the same as in \cite{LS}-\cite{FMS} definition of the
lattice angle functions $\T_{\g_{\xb }}({\xb },{\yb })$\ and
$\T_{\d_{\xb }}({\xb },{\yb })$\ that correspond to the two
opposite types of cuts ($\g$-type and $\d$-type), and
the same definition of
ordering of lattice sites ($\xb > \yb \ $ or $\ \xb < \yb $).
The corresponding two types of
{\it disorder operators} $K_i ({\bf x}_{\g })$ and
$K_i ({\bf x}_{\d })$, \ $i=1,...,N$, are introduced in the form
\beq                \label{3}
{\begin{array} {l}
K_j ({\bf x}_{\g }) =
\exp \left ( i\nu \sum_{\yb \neq \xb } \Theta_{\g_\xb
}({\xb }, {\yb })\ c_j^\dg ({\yb }) c_j ({\yb })\right ),

\vspace{1mm}\\
K_j ({\bf x}_{\d }) =
\exp \left ( i\nu \sum_{\yb \neq \xb }
\Theta_{\d_\xb }({\xb }, {\yb })
\ c_j^\dg ({\yb }) c_j ({\yb })\right ).
\end{array}   }
\eeq
Anyonic oscillators (AOs)\ $a_i(\xb_{\g })$ and
$a_i(\xb_{\d })$,\ \ $i=1,...,N$,  are defined \cite{LS} as
\beq                         \label{4}
a_i(\xb_{\g }) = K_i ({\bf x}_{\g })\ c_i (\xb ) , \qquad\qquad
a_i(\xb_{\d }) = K_i ({\bf x}_{\d })\ c_i (\xb )
\eeq
(no summation over $i$), and the number $\nu $ appearing in (3)--(4)
is usually called the {\it statistics parameter}.
 One can show that these AOs
satisfy the following relations of permutation. For $i\neq j$ and
arbitrary $\xb ,\yb \in \O $,
\beq                              \label{5}
\{ a_i (\xb_{\g }), a_j (\yb_{\g })\} =
\{ a_i (\xb_{\g }), a_j^\dg (\yb_{\g })\}=0.
\eeq
With $q\equiv\exp (i\pi\nu )$, for $i=j$ and for two distinct sites
(i.e. $\xb\ne \yb$) on the lattice $\O$ one has
\beq  a_i (\xb_{\g })                   \label{6}
a_i (\yb_{\g }) +q^{-{\rm sgn }(\xb -\yb )}
a_i (\yb_{\g }) a_i (\xb_{\g }) = 0,
\eeq
\beq                                       \label{7}
 a_i (\xb_{\g })a_i^\dg (\yb_{\g }) +q^{{\rm sgn}(\xb -
\yb )}\ a_i^\dg (\yb_{\g }) a_i (\xb_{\g }) = 0,
\eeq
whereas on the same site
\beq                                       \label{8}
\bigl (a_i (\xb_{\g })~\bigr )^2 = 0,
\qquad\quad\qquad \{ a_i (\xb_{\g }), a_i^\dg (\xb_{\g })\} = 1.
\eeq
The analogs of relations (5)--(8) for anyonic oscillators
of the opposite type $\d$ are obtained by replacing
$\g\to\d$ and $q\to q^{-1}$ in (5)--(8). Clearly, at $\nu =0$
(i.e., $q=1$) anyonic operators reduce to the above fermionic ones.

Note that it is the pair of relations (6), (7) (their analogs for
the $\d$-type of cut, and Hermitian conjugates of all them) which the
statistics parameter $\nu$ does enter. In comparison with ordinary
fermions (1)--(2), the basic feature of anyons is their nonlocality (the
attributed cut) and their peculiar braiding property
encoded in (6), (7). These relations imply that anyons {\it of the
same sort}, even allocated at different sites of the lattice,
nevertheless 'feel' each other due to the factor involving
the parameter $q$ (or $\nu$).

Finally, the commutation relations for anyons of opposite types of
non-locality, i.e. of $\g$-type and $\d$-type, are to be exhibited
($\xb$, $\yb$ arbitrary):
\beq                                              \label{9}
\{ a_i (\xb_{\g }), a_j (\yb_{\d })\} = 0,
\qquad\qquad\quad
\{ a_i (\xb_{\g }), a_j^\dg (\yb_{\d })\}=0
 \ \ \ (\xb \ne \yb),
\eeq
\beq                                                \label{10}
\{ a_i (\xb_{\g }), a_j^\dg (\xb_{\d })\} = \d_{ij}
\ q^{\left [\sum_{\yb <\xb } -
\sum_{\yb >\xb }\right ] c_i^\dg (\yb )c_i (\yb )},
\eeq
as well as the relations which result from (5)--(10) by
applying Hermitian conjugation.

\vspace {4mm}
\noindent  {\bf 3. The algebra $U_q(sl_N)$ and its anyonic realization}

\vspace {3mm}
We adopt the standard notation
$[A]_q\equiv (q^A -q^{-A})/(q-q^{-1})\ $ for $A$ being either
an operator or a number, with a complex number $q$.
Let $a_{ij}$,\ $i,j=1,...,N-1$, be the Cartan matrix of
the $sl_N$ algebras (i.e. $a_{ii}=2$,
\ $a_{i,i+1}=a_{i+1,i}=-1$,\ all the other $a_{ij}$ equal zero).
Quantum algebra $U_q(sl_N)$ is generated by
the Chevalley-basis elements $E_i^{\pm},\ H_i,$\ $i=1,..,N-1$,
which obey the commutation relations
\beq                                         \label {11}
{ \begin{array} {l}
 \qquad\qquad[ H_i, H_j ] = 0 ,

\vspace{2mm}\\

 \qquad\qquad[ H_i, E_j^{\pm} ] = \pm\ a_{ij}\ E_j^{\pm} ,

\vspace{2mm}\\

 \qquad\qquad[ E_i^+, E_j^- ] = \d_{ij}\ [H_i]_q ,

\vspace{2mm}\\

 \qquad\qquad[ E_i^{\pm}, E_j^{\pm} ] = 0 \qquad {\rm if} \quad
                       \vert i-j\vert >1 ,
\vspace{3mm}\\

\hspace{12mm}
(E_i^{\pm})^2\ E_{i+1}^{\pm}-[2]_q E_i^{\pm} E_{i+1}^{\pm} E_i^{\pm}
+ E_{i+1}^{\pm} (E_i^{\pm})^2 = 0,

\vspace{3mm}\\

\hspace{12mm}
 (E_{i+1}^{\pm})^2\ E_i^{\pm}
        - [2]_q E_{i+1}^{\pm} E_i^{\pm} E_{i+1}^{\pm}
        + E_i^{\pm} (E_{i+1}^{\pm})^2 = 0 .
\end{array}   }
\eeq
The $q$-deformed algebra $U_q(sl_N)$ becomes
a Hopf algebra if one imposes
a comultiplication $\D$, counit $\e$, and antipode $S$
(we omit them since they will not be used below).

In what follows, we use the fact of principal importance: the
fundamental representations of the algebras $U_q(sl_N)$ and $sl_N$
are the same, as can be seen from explicit formulas
for representation operators [2].
By repeated use of comultiplication applied to the fundamental
representation, all other representations can be obtained.
Due to different comultiplication rules in these two algebras,
the difference between $U_q(sl_N)$ and $sl_N$ emerges in all
representations other than fundamental and trivial ones.

So, using the lattice of ordered points $\O$, we assign to each
its point $\xb\in\O$ a fundamental $N$-dimensional representation
of the algebra $U_q(sl_N)$.

The following statement is true \cite{CM}-\cite{FMS}.

\noindent{\bf Proposition.}
The set of $N$ anyons defined according
to (4), through the formulae
\beq                                  \label {12}
\begin{array} {l}
E_j^+\equiv A_{j,j+1} = \sum_{x\in\O} A_{j,j+1}(\xb ),  \qquad\qquad
E_j^-\equiv A_{j+1,j} = \sum_{x\in\O} A_{j+1,j}(\xb ),

\vspace{2mm} \\
H_j\equiv A_{jj}-A_{j+1,j+1} =
\sum_{x\in\O}\{ A_{jj}(\xb )-A_{j+1,j+1}(\xb )\}
\end{array}
\eeq
where
\beq                                             \label{13}
\begin{array} {l}
      A_{j,j+1}(\xb ) = a_j^\dg (\xb_{\g}) a_{j+1}(\xb_{\g}),
\qquad\qquad
A_{j+1,j}(\xb ) = a_{j+1}^\dg (\xb_{\d }) a_j (\xb_{\d }),

\vspace{2mm}\\
A_{jj}(\xb ) = a_j^\dg (\xb_{\g}) a_{j}(\xb_{\g}) =
a_j^\dg (\xb_{\d}) a_{j}(\xb_{\d}) =
N_j (\xb ), \ \ \ \ \ \
N_j (\xb ) \equiv c_j^\dg (\xb ) c_{j}(\xb ),
\end{array}
\eeq
provides a (bilinear) realization of the $U_q(sl_N)$ algebra.

These same $N$ anyons, through the formulae
\[
\begin{array} {l}
\tilde{E}_j^+\equiv \tilde{A}_{j,j+1} =
\sum_{x\in\O} \tilde{A}_{j,j+1}(\xb ),
  \qquad\qquad
\tilde{E}_j^-\equiv \tilde{A}_{j+1,j} =
 \sum_{x\in\O} \tilde{A}_{j+1,j}(\xb ),
\vspace{2mm} \\
\tilde{H}_j\equiv \tilde{A}_{jj}-\tilde{A}_{j+1,j+1} =
\sum_{x\in\O}\{ \tilde{A}_{jj}(\xb )-\tilde{A}_{j+1,j+1}(\xb )\}
\end{array}
\]
where
\[                      
\begin{array} {l}
      \tilde{A}_{j,j+1}(\xb ) = a_j^\dg (\xb_{\d}) a_{j+1}(\xb_{\d}),
\qquad\qquad \tilde{A}_{j+1,j}(\xb ) =
 a_{j+1}^\dg (\xb_{\g }) a_j (\xb_{\g }),
\vspace{2mm}\\
\tilde{A}_{jj}(\xb ) = a_j^\dg (
\xb_{\a}) a_{j}(\xb_{\a}) = N_j (\xb )
\end{array}
\]
($\a$ denotes either $\g$ or $\d$), provide
the so-called {\it dual} bilinear realization of $U_q(sl_N)$.
It is easy to see that the generators $\tilde{A}_{jj}$ and
$\tilde{H}_{j}$ in the dual realization coincide with their
counterparts ${A}_{jj}$ and ${H}_{j}$ from the realization (12), (13).

\vspace {4mm}
\noindent {\bf 4. State vectors of decuplet baryons within
anyonic realization}

\vspace {3mm}
In order to construct state vectors for baryons ${\frac32}^+$
that form the decuplet of $U_q(su_3)$,
we exploit the chain of embeddings
\[
U_q(su_3) \subset U_q(su_4) \subset U_q(su_5)
\]
and the respective chain of embeddings of representation spaces
for $[30]\subset[300]\subset[4000]$.
Whereas in the case of adjoint representations the construction
(within anyonic realization) of basis vectors in
their carrier spaces involves  only two sites of lattice,
the baryonic case at hand requires four sites to be involved
in the construction of states from the representation space of
$[4000].$ For definiteness, let us fix the order
${\bf x}_1<{\bf x}_2<{\bf x}_3<{\bf x}_4$ of the coordinates
of the sites.

With the notation (note that each $n_i$ labels the species of anyons
while the subscript '$i$' labels the site of the lattice)
\beq                                               \label{14}
|n_1 n_2 n_3 n_4\ra \equiv  a_{n_1}^\dg(\xb_{1 \d})
a_{n_2}^\dg(\xb_{2 \d}) a_{n_3}^\dg(\xb_{3 \d})
a_{n_4}^\dg(\xb_{4 \d}) \vac ,
\eeq
the highest weight vector (h.w.v.) of $[4000]$ can be realized
as $|1111\ra .$ One can verify this fact merely by checking
that the representation operators  $E_1^+,E_2^+,E_3^+,E_4^+$
annihilate this state,  and the operators
$H_{i},\  i=1,...,4$, acquire on it the appropriate eigenvalues.

Acting on $|1111\ra$ sequentially by the operators
$E_1^-,E_2^-,E_3^-,E_4^-$
we obtain the h.w.v. of representation
$[30]\equiv{\bf {10}}$ (decuplet) space of the subalgebra
$U_q(su_3)$ as certain superposition of states differing
by the position of 5-th sort anyon, and it is natural
to put this into correspondence with the particle
 $\Delta^{++} :$
\beq                                                 \label{15}
|\D^{++}\ra \sim ( |5111\ra + q^{-1} |1511\ra+
q^{-2} |1151\ra+  q^{-3} |1115\ra ).
\eeq
The other basis vectors of ${\bf {10}}$
can be obtained by acting with the lowering operators $E_1^-,E_2^-.$

In order to construct {\it dual} basis (obtainable by acting
with lowering operators in {\it dual} anyonic realization)
we start with the same h.w.v. $|1111\ra$ of the representation
$[4000]$. This fact is reflected in the equality
$|1111\ra=\widetilde{|1111\ra }$, i.e., the h.w.v.
 is common for both bases.

Since a state of real baryon must be insensitive to the choice of cut
type, the state
\beq                                       \label{16}
\widetilde{|\D^{++}\ra } \sim ( q^{-3}|5111\ra + q^{-2} |1511\ra+
q^{-1} |1151\ra +|1115\ra )
\eeq
from dual basis corresponds to the $\Delta^{++}$ hyperon {\it as well}
(let us emphasize that this tilded state serves as a h.w.v.
for irrep ${\bf {10}}$ in dual realization).

We will denote the basis vectors of representation space of
${\bf {10}}$  by $|B_i \ra$ and $\widetilde{|B_i \ra},$
$i=1,...,10, $ where tilded vectors correspond to dual realization.

The normalization for the vectors $|B_i \ra$ and
 $\widetilde{|B_i \ra}$  is chosen in such a way that the condition
\[
\Vert\ |B_i\ra\ \Vert^2 \equiv \widetilde{\la B_i} | B_i \ra=1
\]
holds. This also implies  $\Vert \ \widetilde{|B_i \ra }\ \Vert^2 \equiv
\la B_i \widetilde{|B_i \ra }=1$. Thus, nonlocality is absent in the norm.

For general state vector, we introduce the following
$q$-weighted superposition ($q$-symmetrized state):
\beq                                           \label{17}
|(n_1 n_2 ... n_l)\ra = \sum _{\s \in P_l /
(P_{k_1} \times P_{k_2} \times ... P_{k_r})}{}
|n_{\s(1)}n_{\s(2)}...n_{\s(l)}\ra
q^{\sum_{i=1}^{l}\sum_{j=i+1}^{l}\theta(n_{\s(j)}-n_{\s(i)})} \ \ .
\eeq
In this formula, r is the maximal number from the set
$(n_1,n_2,...,n_l)$, $k_s $($s=1,...,r$) is the multiplicity of
numbers '$s$' in this set, $P_l$ is the group of permutations of
$l$ elements (all the $P_0$ must be omitted), $\theta(x)$
is the usual step function  ( $\theta(x)=1$  if $x>0$ , and
$\theta(x)=0$ otherwise).
All possible permutations of the numbers $(n_1,n_2,...,n_l)$,
which lead to non-coinciding basis vectors, are to be taken
into account in the sum in (17).
As a result, the vector $|(n_1 n_2 ... n_l)\ra$ depends only
on the set of numbers $(n_1,n_2,...,n_l)$ but not on their order.

Using anyonic realization (12)--(13) of the generators $E_i^{\pm}$,
we have the following formulas of their action upon vectors (17):
\beq                                          \label{18}
E_i^+|(n_1 n_2 ...(i+1)...  n_l)\ra=
[k_i+1]_q|(n_1 n_2 ...(i)...  n_l)\ra ,
\eeq
\beq                                          \label{19}
E_i^-|(n_1 n_2 ...(i)...  n_l)\ra=
[k_{i+1}+1]_q|(n_1 n_2 ...(i+1)...  n_l)\ra ,
\eeq
where $k_i$ (resp. $k_{i+1}$) is the multiplicity of $'i'$
(resp. $'i+1'$) in the initial vector.

It is possible to introduce the dual analogue of (17):
\[
\widetilde{|(n_1 n_2 ... n_l)\ra } = \sum _{\s \in P_l /
(P_{k_1} \times P_{k_2} \times ... P_{k_r})}{}
|n_{\s(1)}n_{\s(2)} ... n_{\s(l)}\ra
q^{\sum_{i=1}^{l}\sum_{j=i+1}^{l}\theta(n_{\s(i)}-n_{\s(j)})} .
\]
The dual analogs of the formulas (18) and (19)
(obtained by changing the generators and vectors
into dual ones) are also valid.

The norm of the vector (17) as well as of its dual counterpart is
\beq                                                \label{20}
||\ |(n_1 n_2 ... n_l)\ra\ ||^2= ||
\ \widetilde{|(n_1 n_2 ... n_l)\ra}\ ||^2=
\widetilde{\la(n_1 n_2 ... n_l)}|(n_1 n_2 ... n_l)\ra=
[l]_q!\prod_{s=1}^r\frac{1}{[k_s]_q!}
\eeq
where the notation
$[m]_q!\equiv [m]_q[m-1]_q[m-2]_q\ldots [2]_q[1]_q$
for $q$-factorial $[m]_q$ is used.

Now, in terms of vectors (17) (i.e., the $q$-weighted superposition)
we have the following normalized baryon states :
\beq{                                              \label{21}
\begin{array}{l}
|\D^{++}\ra=\frac{1}{ \sqrt{[4]_q} }|(1115)\ra\ ,\ \ \
|\Sigma^{*+}\ra=\frac{1}{ \sqrt{[3]_q[4]_q} }|(1135)\ra\ ,

\vspace{2mm}  \\
|\Xi^{*0}\ra=\frac{1}{ \sqrt{[3]_q[4]_q} }|(1335)\ra\ ,\ \ \
|\O^{-}\ra=\frac{1}{ \sqrt{[4]_q} }|(3335)\ra .
\end{array} }
\eeq
Thus, we have constructed the state vectors for isoplet
representatives of baryons from ${\bf\underline{10}}.$


\vspace {4mm}
\noindent {\bf 5. Masses of decuplet baryons from anyonic
realization}

\vspace {3mm}
Mass $M_{B_i}$ of a particle $B_i$ is defined as
diagonal matrix element
\beq                                                   \label{22}
M_{B_i}=\widetilde{\la B_i}|\hat{M}|B_i\ra 
\eeq
of the mass operator (see [14])
\[
\hat{M}=M_0 {\bf 1}+\a ( A_{35} \tilde{A}_{53} + \tilde{A}_{35} A_{53} )
 + \b ( A_{53} \tilde{A}_{35} + \tilde{A}_{53} A_{35} ) \ ,
\]
where $A_{35}\equiv A_{35}(q)\equiv q^{1/2} A_{34} A_{45}
- q^{-1/2} A_{45} A_{34},\ \ $
 $A_{53}\equiv A_{53}(q)\equiv q^{1/2} A_{43} A_{54}
- q^{-1/2} A_{54} A_{43},\ \ $
$\tilde{A}_{35}\equiv -A_{35}(q^{-1}),\ \ $
$\tilde{A}_{53}\equiv -A_{53}(q^{-1})$.
Clearly, the constants $M_0$, $\a$ and $\b$ have the dimension
of mass.

Due to particular choice of representations (mentioned above),
the part of mass operator which gives nonvanishing contribution
becomes
\[
\hat{M}=M_0 {\bf 1}+\a \hat{M}_\a + \b \hat{M}_\b
\]
where
$\ \hat{M}_\a \equiv E_3^+E_4^+E_4^-E_3^-, \ $
$\hat{M}_\b \equiv E_4^-E_3^-E_3^+E_4^+. \ $
With the notation
\[
M_\a(B_i) \equiv\widetilde{\la B_i }|\hat{M}_\a| B_i \ra,
\ \ \ \ \
M_\b(B_i) \equiv \widetilde{\la B_i }|\hat{M}_\b| B_i \ra,
\]
the expressions for masses take the form
\beq
M_{B_i}=M_0 +\a M_\a(B_i)  +\b M_\b(B_i).
\eeq
Then, using the relations $(E^{\pm}_i)^*=\tilde{E}_i^{\mp}$
where $"*"$ denotes Hermitian conjugation, one rewrites
the expressions $M_\a(B_i)$ and $M_\b(B_i)$ as follows:
\beq
M_\a(B_i) \equiv || E_4^-E_3^- | B_i \ra ||^2,\ \ \ \ \
M_\b(B_i) \equiv || E_3^+E_4^+ | B_i \ra ||^2.
\eeq
In other words, $M_\a(B_i)$ is given as scalar product of
the vectors $E_4^-E_3^- | B_i \ra$ and their duals
$\tilde{E}_4^-\tilde{E}_3^- | \widetilde{B_i \ra };$ likewise,
$M_\b(B_i)$ is given by the scalar product of vectors
$E_3^+E_4^+ | B_i \ra$ and their duals
$\tilde{E}_3^+\tilde{E}_4^+ | \widetilde{B_i \ra }.$

It is not hard to see that, with this specific form of
mass operator, masses of baryons within each isomultiplet
(multiplet of $U_q(su_2)$) in the decuplet are equal.

In order to obtain the expressions for masses of baryons
with state vectors (21), we substitute these vectors
in formulas (23),(24) and take into account (18)--(20).
For example, for the $\O^-$ hyperon we obtain
\[ M_{\O^-} =M_0 + \a M_\a(\O^-) + \b M_\b(\O^-)
\]
with
\[
M_\a(\O^-)=||E_4^- E_3^- |\O^-\ra||^2
=\frac{1}{[4]_q}||E_4^- E_3^- |(3335)\ra||^2=
\frac{[2]_q^2}{[4]_q}||\ |(3355)\ra\ ||^2
=[2]_q[3]_q
\]
(here the formulae $E_3^-|(3335)\ra=|(3345)\ra$,
$E_4^-|(3345)\ra=[2]_q|(3355)\ra $, and (20) were used)
and
\[
M_\b(\O^-)=||E_3^+ E_4^+ |\O^-\ra||^2
=\frac{1}{[4]_q}||E_3^+ E_4^+ |(3335)\ra||^2=
\frac{[4]_q^2}{[4]_q}||\ |(3333)\ra\ ||^2=[4]_q
\]
(here the formulae
$E_4^+|(3335)\ra=|(3334)\ra$, $E_3^+|(3334)\ra=[4]_q|(3333)\ra $,
and (20) were used).      Hence,
\beq
M_{\O^-} =M_0 + [2]_q[3]_q \a  +  [4]_q \b .
\eeq
In similar way, calculations for the other three isoplets yield the masses
\beq
M_{\D}=M_0+\b,\ \ \ \ \ \
M_{\Sigma^{*}}=M_0+[2]_q\a + [2]_q\b,\ \ \ \ \ \
M_{\Xi^{*}}=M_0 + [2]_q^2\a + [3]_q\b.
\eeq
Eliminating the constants
$M_0\ ,\a\ ,\b\ $ from the expressions (25),(26)
for isoplet masses of the decuplet, we arrive at the mass relation
\beq
{M_{\O^-}-M_{\Xi^*}+M_{\Sigma^*}-M_{\D}}
=
[2]_q\ (M_{\Xi^{*}}-M_{\Sigma^{*}})
\eeq
in the form of $q$-average. It is the same mass relation as that
obtained previously in \cite{GKT} using $q$-analogue of
Gel'fand-Tsetlin formalism.
In \cite{GKT} it was also shown that the decuplet mass formula
(27) results not only within this particular dynamical  $U_q(su_5)$
representation $[4000],$ but also within any admissible (such
that contains the $U_q(su_4)$ ${\sl 20}$-plet wherein the decuplet
of $U_q(su_3)$ is placed) dynamical representation of $U_q(su_5)$ or
 $U_q(su_{4,1})$. This is
{\it a kind of universality} - independence of the result (27)
on the choice of dynamical representation.

Comparison of (27) with empirical situation [15]
shows that with $q=\exp ({\rm i} \pi / 14)\ $
(in which case we have $[2]_q\simeq 1.96$),
the mass relation  remarkably agrees with data.
This is in contrast with the  average-type formula
$\frac12 (M_{\O^-}-M_{\Xi^*}+M_{\Sigma^*}-M_{\D})
= M_{\Xi^{*}}-M_{\Sigma^{*}}$ known long ago [16].
To see that, it is enough to ``predict'' $M_{\D}$,
both from the latter formula and from
the $q$-deformed one (27)
with $[2]_q\simeq 1.96$, in terms of known masses
$M_{\O^-},\ M_{\Xi^*},\ M_{\Sigma^*}$,
and then to compare the results with the empirical value
$(1232\pm 2)$~MeV of $M_{\D}$. Unlike the case of
average-type formula,
the value of $M_{\D}$ ``predicted'' on the base of (27)
lies exactly in the range $(1232\pm 2)$~MeV.
Concerning physical meaning of the value $q=\exp ({\rm i} \pi / 14)\ $
of deformation parameter it was argued in \cite{G97} that
$\frac{\pi}{14}$ may be identified with the Cabibbo angle.

\vspace {4mm}
\noindent {\bf 6. Concluding remarks}

\vspace {3mm}
   We have shown in the decuplet case that
the $U_q(su_N)$ based results \cite{GKT} concerning decuplet
baryon masses and their mass sum rule can as well be derived
in the framework of anyonic realization of the (flavour symmetry)
quantum algebras. To this end, we have
first constructed, using anyonic operators, the state vectors
corresponding to isoplet members of the decuplet. Then we
have calculated baryon masses adopting the appropriate
definition (22) which allowed us to get rid of nonlocality
in final results for the masses.
As a consequence it appears that constituent quarks (as modified
within the present approach) may be, at least formally,
considered as anyons of definite statistics $\nu $.
From comparison of the $q$-relation (27) with empirical data
for decuplet baryons we draw the conclusion that:
the anyonic statictics parameter (connected with the
deformation parameter $q$) should be $\nu ={1\over 14}$ in
the decuplet case. Recall that the value of statistics parameter
characterizes {\it braiding of anyons} (on the lattice $\O$),
and these are used as building blocks
for the state vectors of decuplet baryons, as seen in Sec. 5.
Of course, there still remains
an open question whether the quarks bound in
baryons ${\frac32}^+$ could be really considered
as anyon-like, quasi-two-dimensional entities.
This problem however has to be
answered on the base of a field-theoretic setting which
goes beyond the scope of present note.  Let us only mention,
as partial justification, that one can use the argumentation
of Ref. [17] according to which the appropriately interacting
(charge-flux) fermions, with nonzero probability to be found
in the plane (in which they interact), exhibit
to certain extent the properties of anyonic statistics.

\medskip

We thank Prof. R.~Jackiw for discussion and useful
remarks. The research described in this publication was made possible
in part by the Award No. UP1-309 of ths U.S. Civilian Research
\& Development Foundation for the Independent States of the
Former Soviet Union (CRDF).

\bigskip


\end{document}